\documentclass[10pt,letterpaper]{article}

\usepackage{cogsci}

\cogscifinalcopy 
\usepackage{array}
\usepackage{siunitx}
\usepackage{pslatex}
\usepackage[natbibapa]{apacite}
\usepackage{graphicx}
\usepackage{enumitem}
\usepackage[bottom]{footmisc}
\usepackage{subcaption}
\usepackage{times}
\usepackage{latexsym}
\usepackage{amsmath}
\usepackage{algorithm}
\usepackage{algorithmic}
\usepackage{graphicx}
\usepackage{subcaption}
\usepackage{array}
\usepackage{caption}
\usepackage{tabularx}
\usepackage[table,xcdraw]{xcolor}
\usepackage{float}
\usepackage{amsmath}
\usepackage{url}
\usepackage{placeins}
\graphicspath{{figures/}}
\usepackage{hyperref}
\hypersetup{
    colorlinks,
    citecolor=blue!50!black,
    filecolor=black,
    linkcolor=black,
    urlcolor={blue!80!black}
}
\usepackage{amssymb}
\usepackage{booktabs}
\usepackage{multirow}
\usepackage[group-separator={,}]{siunitx}

\usepackage{xspace,tcolorbox}
\usepackage[dvipsnames]{xcolor}

\makeatletter
\renewcommand\@afterheading{%
  \@nobreaktrue
  \everypar{%
    \if@nobreak
      \@nobreakfalse
      \clubpenalty 1
      \if@afterindent \else
        {\setbox\z@\lastbox}%
      \fi
    \else
      \clubpenalty 1
      \everypar{}%
    \fi}}
\makeatother

\makeatletter
\renewenvironment{quote}
  {\list{}{\listparindent=0em
           \itemindent=\listparindent
           \leftmargin=\parindent
           \rightmargin=\leftmargin
           \topsep=0em
           \parsep\z@\@plus\p@}
   \item\relax}
  {\endlist}
\makeatother

\usepackage{etoolbox}
\usepackage{environ}

\newtoggle{bibdoi}
\newtoggle{biburl}
\makeatletter

\undef{\APACrefURL}
\undef{\endAPACrefURL}
\undef{\APACrefDOI}
\undef{\endAPACrefDOI}

\long\def\collect@url#1{\global\def\bib@url{#1}}
\long\def\collect@doi#1{\global\def\bib@doi{#1}}
\newenvironment{APACrefURL}{\global\toggletrue{biburl}\Collect@Body\collect@url}{\unskip\unskip}
\newenvironment{APACrefDOI}{\global\toggletrue{bibdoi}\Collect@Body\collect@doi}{}

\title{Cognitive offloading and the speedup illusion in human-AI interaction}

\author{
Sunny Yu$^{1,2}$, Myra Cheng$^{1}$, Ahmad Jabbar$^{3}$, 
Ilia Sucholutsky$^{4}$, Katherine M. Collins$^{5,6}$, 
Dan Jurafsky$^{1,3}$, Robert D. Hawkins$^{3}$ \\
\texttt{syu03@stanford.edu} \\
$^{1}$Department of Computer Science, Stanford University \\
$^{2}$Symbolic Systems Program, Stanford University \\
$^{3}$Department of Linguistics, Stanford University \\
$^{4}$New York University \\
$^{5}$MIT \\
$^{6}$Princeton AI Lab
}

\begin{document}

\maketitle

\begin{abstract}

Large language models (LLMs) have the potential to boost human productivity by speeding up task completion---provided users know when to offload cognitive work to them. But we do not know if users are well-calibrated in estimating these potential time savings. We conducted a preregistered large-scale behavioral study (N = 1237) to characterize mismatches between expectations and reality, with a focus on simple cognitive tasks. While actual completion times between independent completion and AI-assisted completion did not differ, participants predicted AI to be significantly faster. The same bias was not observed when imagining help from another human participant. We identify a \emph{speedup illusion} where people have accurate forecasts of independent completion times but significantly underestimate AI-assisted times. Additionally, time and effort dissociate: participants reported lower subjective effort with AI despite equivalent completion times. This suggests that completion time itself is not sufficient to characterize efficiency gains.

\end{abstract}

\section{Introduction}

People routinely offload cognition to external resources: writing things down, using calculators, consulting others, or searching the web \citep{fan2023drawing}. From a resource-rational perspective, such offloading reflects adaptive allocation of limited cognitive resources \citep{lieder2020resource, griffiths2019doing}. 
The decision of whether to offload a given task depends on a cost-benefit comparison: people should delegate to external tools when internal costs exceed the costs of using external support \citep{risko2016cognitive}. But this comparison requires \emph{calibration}, an accurate mental model of both one's own capabilities and those of the external tool.

For many cognitive tools, people appear reasonably calibrated: they offload more when internal memory is taxed and when tasks are difficult \citep{dunn2016toward, wahn2023offloading}. Large language models (LLMs) represent a new class of cognitive tool, one that can serve as a ``thought partner'' \citep{collins2024building} with broader and more variable capabilities than calculators or search engines \citep{xiong2024search, hooper2025cognitive}. 
The prevailing narrative suggests these tools dramatically boost productivity \citep{handa2025economic, wang2025ai, anthropic2026aeiv4}.

While AI assistance offers efficiency gain on complex tasks that would otherwise take a long time for humans to complete independently, it is unclear whether using AI can save time on simpler problems. Metacognitive monitoring provides us with imperfect but usable signals about our own cognitive processes \citep[how long a task will take, how difficult it feels, whether we're on track;][]{koriat2015metacognition}. We lack such privileged access to the internal workings of an LLM. This asymmetry suggests people may be systematically biased in their predictions of AI-assisted completion times compared to their independent completion times: a \emph{speedup illusion}. Emerging evidence is consistent with this concern: \citet{becker2025measuring} found that AI assistance slowed down coding by 19\% despite an expectation for speedup. Other studies show mixed results, with some finding AI reduces subjective effort \citep{stadler2024cognitive} and others finding no effect \citep{dhillon2024shaping}.

Are people well-calibrated about how much time using AI can save? More specifically, can people identify that AI assistance does not necessarily save time on simple cognitive tasks, or do they overestimate the extent to which AI assistance leads to an efficiency gain? Based on the metacognitive access argument \citep{overgaard2012kinds}, we hypothesize that there is an \emph{asymmetric miscalibration}: people are reasonably calibrated about their own completion times but systematically underestimate the AI-assisted time.

To test the hypothesis, we collected data for predicted and actual completion times for both independent and AI-assisted work. We conducted a pre-registered, large-scale behavioral study ($N = 1,237$)\footnote{This
study was approved by the Stanford Institutional Review Board (IRB) under protocol 83204 and
pre-registered at \href{https://osf.io/8x9j6}{OSF}.
The data and code can be accessed at
\href{https://github.com/sunnyych/cognitive_offloading_cogsci}{GitHub}.} to test the prediction. In the prediction sample, participants predicted how long tasks would take (independently or with assistance); in the completion sample, they actually completed the tasks themselves, either independently or with AI assistance. Tasks spanned four categories of cognitive work, from simple information retrieval to content creation. 
Our results support the asymmetric miscalibration account: participants accurately predicted their independent completion times but significantly underestimated how long tasks would take with AI assistance; participant miscalibration holds across task difficulty. Furthermore, while AI assistance did not reduce actual completion times, it reduced people's experienced subjective effort. Together, these findings suggest that a ``speedup illusion" persists even on easy tasks --- a finding especially worrying if miscalibration encourages AI use, forming a feedback loop where AI use leads to further miscalibration.

\begin{figure*}[t] 
    \centering
    \includegraphics[width=0.9\textwidth]{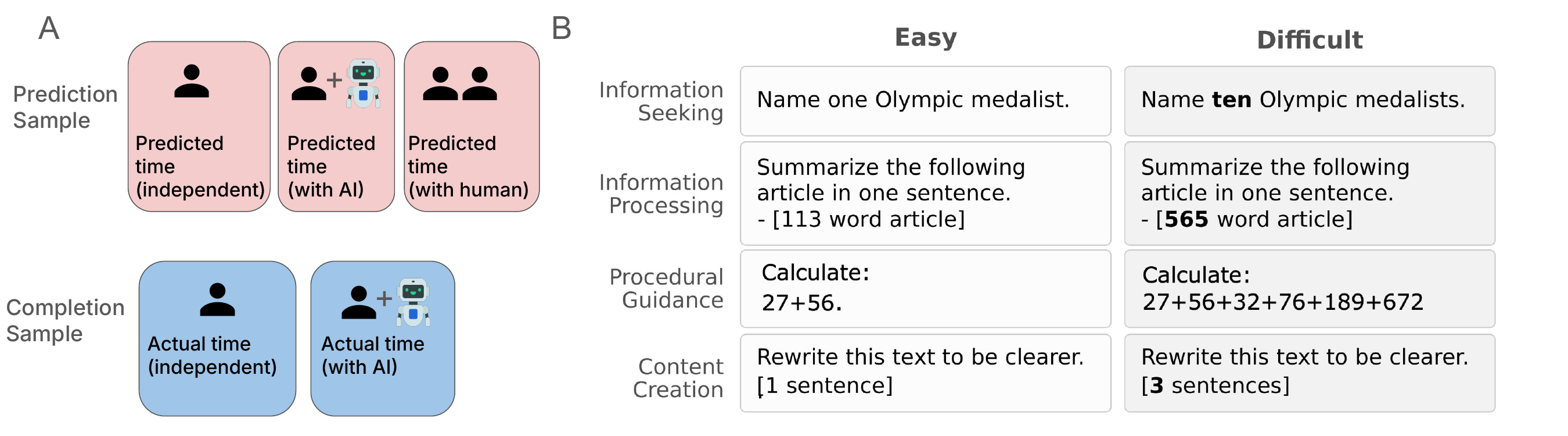} 
    \caption{\textbf{Experiment setup}: \textbf{a)} we include a prediction sample and a completion sample. In the prediction sample, participants predict the completion times for themselves and for using external assistance. In the completion sample, participants complete the task independently or with AI assistance. \textbf{Example tasks}: \textbf{b)}, we show one example per category $\times$ difficulty level (8 of the 24 tasks).}
    \label{fig:fig1}
\end{figure*}

\section{Methods}
\subsection{Problem Formulation}
For a task $\tau$, we notate the time for a human to complete the task independently as $t_H(\tau)$ and the time for a human to complete the task with AI assistance as $t_A(\tau)$.
In addition to the actual completion times, people's expectation for how long a task would take to complete may vary w.r.t. whether the task is being completed independently or with AI assistance. Call a human \textit{well-calibrated} w.r.t. task completion time if they are accurate in their prediction of the completion time of the task. 
We can compare the predicted and actual $t_H(\tau)$ and  $t_A(\tau)$ to ascertain whether people are well-calibrated about their independent and AI-assisted completion times, respectively. The direction and difference between $t_H(\tau)$ and $t_A(\tau)$ reveal whether AI assistance saves time. Similarly, replacing time with self-reported subjective effort, we can examine whether using AI reduces the subjective cognitive effort, thereby exploring whether time aligns with subjective effort.

\subsection{Task Construction}

We constructed a total of 24 tasks (see Figure \ref{fig:fig1}) based on a taxonomy of LLM use, capturing four broad categories of cognitive skills required by different tasks: C1 -- Information Seeking, C2 -- Information Processing \& Synthesis, C3 -- Procedural Guidance \& Execution, and C4 -- Content Creation \& Transformation \citep{shelby2025taxonomy}. We included a total of six tasks for each category, reflecting distinct task types in their taxonomy. The tasks spanned two difficulty levels, differing in the amount of cognitive effort required (e.g., ``Name one Olympic medalist.'' vs. ``Name ten Olympic medalists.'').

\subsection{Experiment setup}

To measure people's completion times and how their expectations for them differed from reality, we included a prediction sample and a completion sample. While a within-subject design would provide more direct insights into calibration patterns at an individual level, it would bias the actual completion times and effort if participants were already familiar with the tasks before completing them. A between-subject design, on the other hand, minimizes downstream biases on the variables and provides insights about average differences.

\noindent
\textbf{Prediction sample.} In the prediction sample, participants were randomly assigned to one of two hypothetical conditions. Without having to complete the task while having a detailed task description, a participant was asked to predict how long it would take to complete the task independently and with AI assistance or with assistance from another human participant. For each task, participants also stated whether they would choose to complete the task independently or with the external assistance of AI/human. After making predictions, participants completed the Need for Cognition scale \citep{cacioppo1982need}. Participants were instructed to not use any form of AI, and copying and pasting was disabled.

\textbf{Completion sample.} In the completion sample, participants were randomly assigned to complete tasks either independently or with AI assistance. For the AI assistance condition, we provided an embedded chat interface with GPT-4o on the task page that participants could interact with. Participants were presented a randomized mix of easy and difficult tasks. Each task was completed by around 68 participants for each condition. We used a hidden timer to record the completion time for each task. After the completion of each task, participants answered the 5-question version of the NASA-TLX, a measure of subjective effort \citep{Hart_1986}. The questions evaluate how mentally demanding (Q1), hurried or rushed (Q2), and successful (Q3) the participant felt completing the task, as well as how hard the participant had to work (Q4), and how insecure, discouraged, and stressed they felt (Q5) when completing the task. To understand people's calibration and efficiency gain when it does not affect the outcome, we filtered out incorrect responses and focused on correct answers only and analyzed the corresponding time and effort on these tasks. To exclude low-effort responses (which would correspond to shorter completion times), we annotated all final answers and excluded ones that were incorrect (for questions with verifiable answers), low-effort, or failed to address the prompts to ensure that all responses were high-quality. In general, participants made a decent effort to complete the tasks across both of the conditions; we excluded 6.3\% of the responses in the independent condition and 4.0\% in the AI assistance condition.

\textbf{Participants.} Participants were recruited through the Prolific platform, comprising a representative sample of the US adult population. Our sample size was N=1237 in total: 401 in the prediction sample and 836 in the completion sample. The participants were 48\% female, 47\% male, and 5\% other; 58\% white, 11\% Black, 10\% mixed, and 6\% Asian.

\section{Results} 

\paragraph{People expect offloading to AI to be more efficient.}
Before examining the calibration gaps, we first confirm that people generally expect AI assistance to reduce task completion times. We fit a linear mixed-effects model with prediction target (independent vs. assisted) and assistance source (AI vs. another participant) as fixed effects, including participant and task-level random intercepts.

Our model reveals that people expect AI assistance to reduce task completion times by 68.5 seconds ($\beta=68.5$, $SE = 3.37$, $z = 20.34$, $p<0.001$). To determine whether the difference observed above is unique to AI assistance, we also ask them about ``another participant'' who is highly intelligent\footnote{Without an anchor, participants' predictions for ``another participant" would conflate two separate beliefs: the average competence of an unknown  other and the predicted speedup given that competence. Anchoring on ``highly intelligent" matches the implicit competence assumption participants likely bring to the AI condition for these tasks (a capable assistant).}) as a baseline. We found that participants estimated independent completion to take significantly longer than completion with the help of another participant ($\beta=17.10$, $SE = 3.36$,  $z = 5.10$, $p<0.001$), although the difference is significantly greater for AI assistance than for another participant ($\beta=-51.4$, $SE = 4.76$,  $z = -10.80$, $p<0.001$), meaning that on average, participants expected offloading to AI to be more efficient than offloading to another participant. At a task level, participants predicted AI assistance to speed up completion times for 18 tasks, and predicted assistance from another participant to speed up 5 of the tasks. The finding is consistent with participants' stated preferences: when asked how they would complete the task, people preferred AI assistance more than assistance from another participant ($\beta=1.37$, $SE=0.26$, $z=5.34$, $p<0.001$). The results show that people expect \textbf{assistance from AI} to be particularly effective in reducing completion times.

\begin{figure}[t] 
    \centering
    \includegraphics[width=\columnwidth]{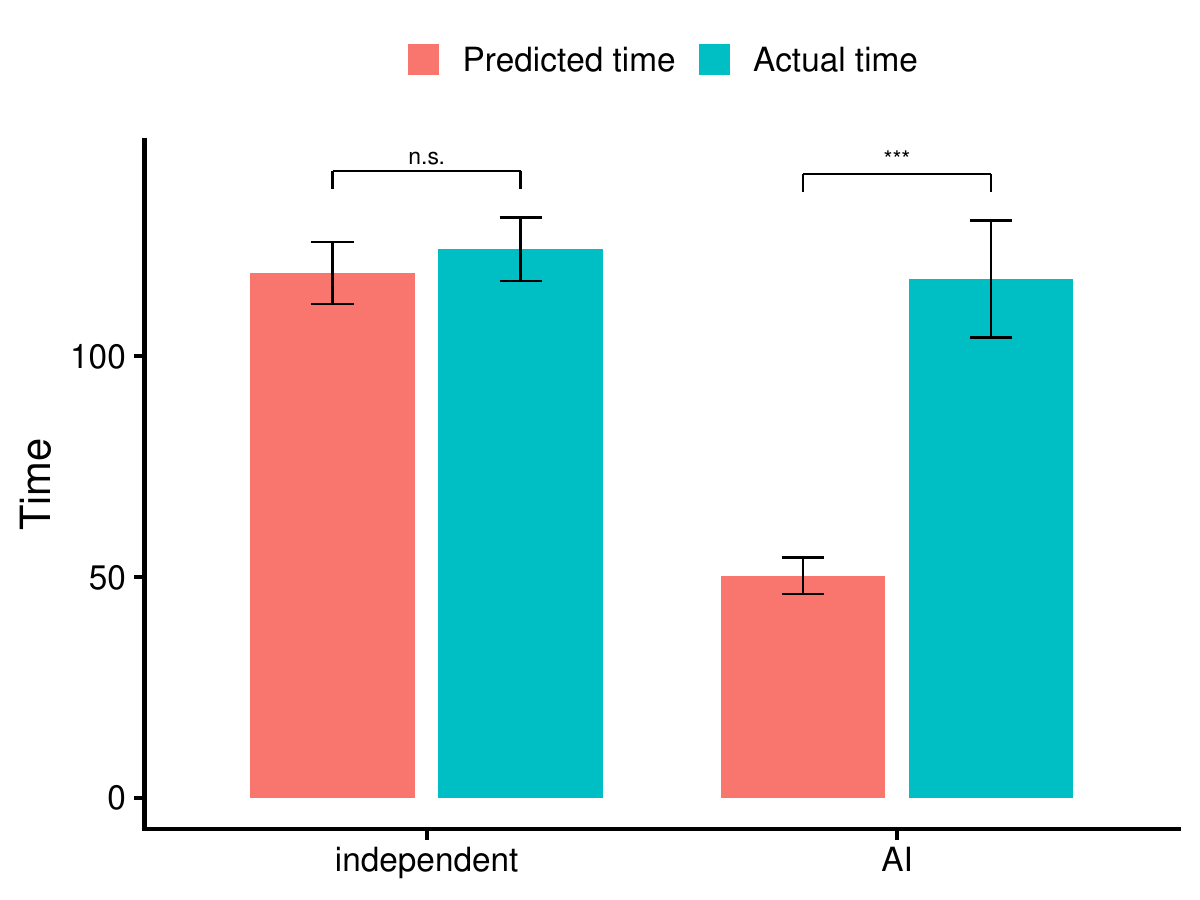} 
    \caption{There is no significant difference between the predicted and actual completion times for the independent completion condition (left), but actual time is significantly greater than predicted time for the AI condition (right).}
    \label{fig:fig2}
\end{figure}

\paragraph{People miscalibrate the time that AI assistance saves.}
To investigate whether people's predictions of completion times with AI assistance is aligned with reality, we fit the following linear mixed-effect model: $\text{time} \sim \text{condition} * \text{type} + \text{difficulty} + (1 \mid \text{participantID}) + (1 \mid \text{taskID})$, where time is the prediction or completion times in seconds, condition is prediction vs. completion, and type is independent vs. AI assistance. We found that people failed to reliably calibrate how much time AI assistance saves (Figure \ref{fig:fig2}): the actual completion time was almost one minute greater than the predicted time ($\beta=57.8$, $SE = 8.79$, $z=6.57$, $p<0.001$). The speedup illusion holds across all four task categories and both difficulty levels.

In stark contrast to the AI assistance condition, we found no significant difference between the actual and predicted completion times in the independent completion condition ($\beta=-2.52$, $SE = 8.83$, $z=-0.29$ , $p=0.775$), which suggests that people's predictions are generally accurate. However, two category-wide differences emerge: participants underestimated their completion time for C2 ($\beta=29.2$, $SE = 11.7$, $z=2.49$, $p<0.05$) but overestimated for C4 ($\beta=-24.25$, $SE = 11.7$, $z=-2.07$, $p<0.05$ ). We observe no significant effect across difficulty levels.

\begin{figure*}[t]
    \centering
    \begin{subfigure}{0.40\textwidth}
        \centering
        \includegraphics[width=\linewidth]{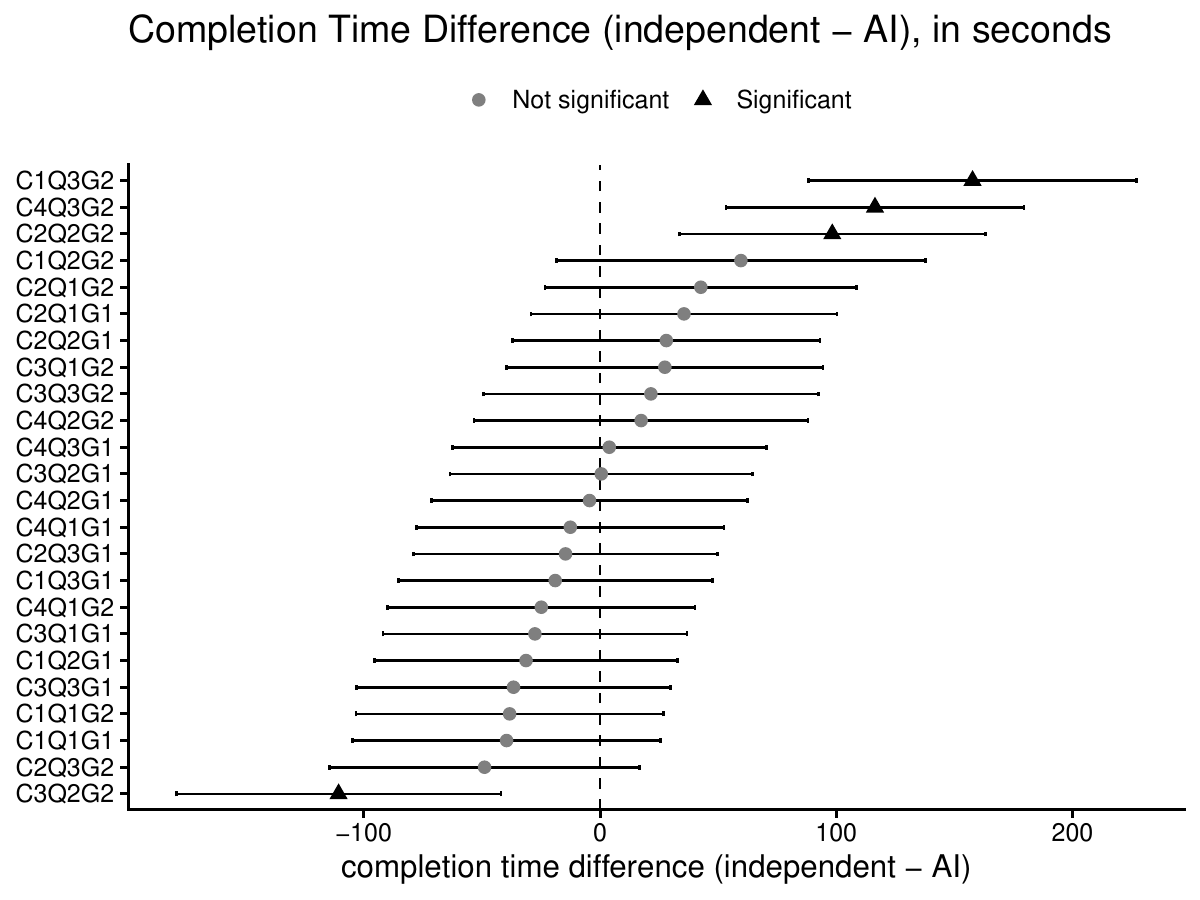}
        \label{fig:diff_time}
    \end{subfigure}\hfill
    \begin{subfigure}{0.40\textwidth}
        \centering
        \includegraphics[width=\linewidth]{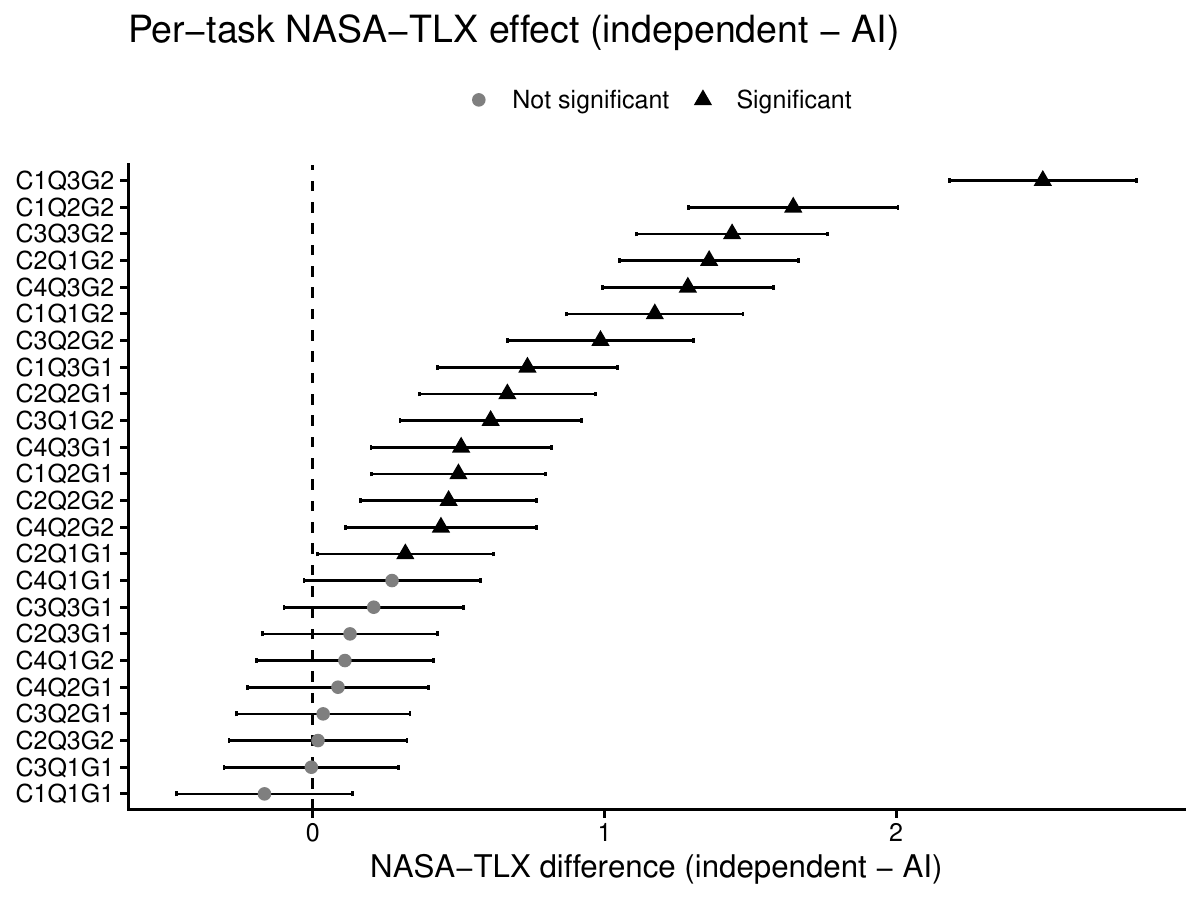}
        \label{fig:diff_nasa}
    \end{subfigure}
    \caption{The difference between the two conditions per task in terms of time (left) and NASA-TLX (right), ordered by effect size. The completion time difference is only significant for 3 tasks. AI assistance leads to offloading (measured by NASA-TLX) for 15 of the tasks (C denotes category, Q denotes question, and G denotes difficulty level, where 1 is easy).
    }
    \label{fig:figure4}
\end{figure*}

We found that while the speedup illusion existed across all tasks, AI assistance only made difficult tasks faster, and not easy ones: task completion with AI assistance was faster by 26.14 seconds ($SE=11.9$, $z=2.19$, $p<0.05$) on difficult tasks, but no difference between independent completion time and AI-assisted completion time was found for easy tasks. Across individual tasks, the completion times only became significantly faster with AI assistance for three tasks (coming up with 10 other words for ``enjoy'', summarizing a long article, and editing a long text; see Figure \ref{fig:figure4}). The results highlight that when it comes to basic cognitive tasks like providing instructions for boiling an egg or finding spelling errors in a sentence, people do not have a good mental model of whether using a resource-rational strategy also makes the completion more efficient. On easy tasks, conserving effort does not translate to a conservation of time.

People's individual traits also influenced their decision to offload cognitive efforts. We examine whether participants' willingness to think (as measured by the Need For Cognition scale) predicts the magnitude of calibration errors. We found that participants who are more averse to thinking were more likely to believe that AI assistance saves more time. Specifically, we focus on the prediction sample and identify negative coefficients for the difference between the predicted independent versus assisted completion times for item 3 (``thinking is not my idea of fun''; $\beta=-6.55$, $SE=3.14$, $t=-2.11$, $p<0.05$) and item 4 (``I would rather do something that requires little thought than something that is sure to challenge my thinking abilities''; $\beta=-6.66$, $SE=3.19$, $z=-2.049$, $p<0.05$) of the NFC scale. This means that the more participants dislike having to think, the more susceptible they are to the speedup illusion, believing that using AI can be significantly faster than doing the task on their own. The finding is consistent with previous research indicating a stronger effect between perceived difficulty and subjective cognitive effort for individuals with low ability or conscientiousness \citep{yeo2008subjective}. On the other hand, variables related to people's familiarity and experience with AI --- namely AI-use frequency and the AI Assessment scale (AIAS) \citep{grassini2023development} --- do not predict the magnitude of calibration error. 

\paragraph{Subjective mental effort differs from time.}

While cognitive offloading is typically measured with time \citep{handa2025economic, wang2025ai}, it is also related to the \textit{experience} of mental effort, which in essence is subjective \citep{steele2020perception}. Bearing in mind that the time taken may not capture the subjective experience of mental effort, we sought to directly record people's mental effort using NASA-TLX index. Our results suggest that subjective mental effort could be a possible reason for the calibration error. Overall, there was a positive and weak correlation between completion time and the NASA-TLX average ($r=0.26$). However, the two measurements differ fundamentally: while AI assistance sped up completion of only a few tasks, it reduced mental effort for all tasks --- on average, the NASA-TLX average (higher means more effortful) lowered by 0.61 points on a 7-point scale ($SE=0.059$, $t=-10.38$, $p<0.001$). 

The significant reduction in perceived subjective effort holds for 15 out of 24 tasks. Comparing the effort in the independent and AI conditions provides us with an informative understanding of the reduction of mental effort with AI use. An AI-assisted reduction in mental effort across all easy and difficult tasks may have contributed to an overestimation of the time saved with AI assistance. This provides us with an understanding of a possible source of the miscalibration error.

\paragraph{How are people prompting AI?}

To understand the different ways that participants prompted the LLM, we provide an in-depth analysis of prompting behaviors and user-LLM interactions in the experiment. More than 70\% of the interactions were single-turn, and the maximum number of turns was 5 for all tasks. Out of all the prompts, 18.5\% were directly copied and pasted from the instructions. Copying and pasting did not significantly reduce completion time, but it reduced NASA-TLX (average) by 0.14 points ($SE=0.058$, $z=2.38$, $p<0.05$). The finding reveals that when completing a task with AI assistance, much of the cognitive effort comes from writing the prompt. 

To verify if prompting took up the majority of the task completion time, we decomposed the completion time into three components: the time it takes to construct the prompt, the time for the model to generate a response, and the time the user reads and processes the model response, i.e., $t_A(\tau) = t_P(\tau) +t_M(\tau) + t_R(\tau)$\footnote{The prompting time is calculated by taking the difference in time stamps between the first model message and the last user prompt. The model response generation time is calculated as the difference in the time stamp between the user's last prompt and the model's final response. Finally, the post-response time is calculated as the difference between the task completion time (the variable we use for all analyses above) and the time between the first model message and the last model message.}. Overall, the time it took for the model to generate a response was minimal (2.89 seconds), so we compare prompting with post-response processing.  Across all tasks, we found no significant difference between prompting and response processing time. But we found evidence that prompting could take significantly longer for some tasks, while taking less time for others: prompting took 33.4 seconds longer ($SE=12.1$, $z=2.76$, $p<0.01$) than post-response processing for C2 tasks. It is possible that people's increased familiarity with AI systems and previous prompting history made it easier and faster for them to write prompts \citep{korzynski2023artificial}.

On the other hand, we found that post-response processing took 113.9 seconds longer than prompting ($SE=29.3$, $z=-3.89$, $p<0.001$) on the logic problem. An analysis of the conversation histories for this question reveals that GPT-4o tended to produce verbose answers that are difficult to understand (see example below). For the same question, AI-assisted completion times were significantly longer ($\beta=-110.66$, $SE=35.1$, $z=-3.16$, $p<0.01$), which contradicted with people's prediction that AI assistance should speed up the task by more than 2 minutes ($\beta=136.31$, $SE=14.2$, $z=9.59$, $p<0.001$). The question highlights a case where people need to spend more time and effort reading and understanding a model's response.

\begin{quote}
\textbf{Task:} If it is raining and it is sunny, then there is a rainbow. It is not the case that there is a rainbow. Therefore, either it is not the case that it is raining or it is not the case that it is sunny. Construct a similar argument with the statements: ``There is water in the pot'', ``The pot is on the stove'', ``The water gets heated. \\
\textbf{Assistant:} ...This argument follows a similar logical pattern to the original, using the form of modus tollens:

- If \( P \land Q \), then \( R \).
- Not \( R \).
- Therefore, not \( P \) or not \( Q \).

Where:
- \( P \): ``There is water in the pot.''
- \( Q \): ``The pot is on the stove.''
- \( R \): ``The water gets heated.''...

\textbf{User:} Umm—this is getting a bit abstruse. Can we summarize this
proposition in simpler terms?
\end{quote}

While copying the tasks directly as prompts was common, we also observed instances where participants wrote very different prompts even for the same task. For example, for the task on naming Olympic medalists, many participants had a specific category or name in mind and used AI as a recall tool (Table \ref{tab:prompt_list}). Some participants used LLMs to confirm or expand on their existing answers. For the trolley problems, where participants had to justify whether to divert the trolley, some participants asked AI to articulate the reasoning rather than making the decision (e.g., ``Give me three reasons in bullet format of why I should pick the 10 students and three reasons in bullet format of why I should pick the five retired doctors''). When using AI to summarize a satire article, some participants chose to have the AI expand on their own summary of the article rather than delegating the AI to do the task entirely, e.g., 
``[The article is] too long to paste, but it's about the many rules and regulations that airlines have that can be burdensome and frustrating.'' In these cases, the participants already expended the majority of the effort required to complete the task prior to receiving the AI answer. The decision to do so may reflect a resource-rational choice to conserve both time and effort. 

Finally, participants actively engaged with the model responses and proactively asked questions like ``Why did you leave out the university's struggles to impliment policy regarding AI'' for the article summarization question or ``Why do you have the temperature so high in the beginning?'' for the explanation question. The wide range of behaviors showcase how differently participants prompted the model and their different degrees of cognitive engagement.

\begin{table*}[t]
\small
\centering
\begin{tabular}{p{0.95\textwidth}}
\toprule
\textbf{Prompt} \\
\midrule
I would like to know who won gold in the 400m men's freestyle in the 2016 Olympics. \\

Tell me who won the women's snowboarding gold medal in the last Winter Olympics. \\

Can you name the Olympic medalist with the highest number of medals? \\

NAME AN OLYMPIC MEDALIST IN SKIING. \\

Please provide a list of men and women 10 gold medalists in the 100m freestyle in the last 10 Olympics. \\

\bottomrule
\end{tabular}
\caption{Examples of participants' prompts to AI for the ``name an Olympic medalist'' task that were not directly copied.}
\label{tab:prompt_list}
\end{table*}

\section{Discussion}
\paragraph{People conflate effort reduction with time reduction.}
Decades of cognitive science research has studied people's biases in the perception and estimation of time, which can be affected by factors like the pleasantness of the experience \citep{fraisse1984perception, fredrickson1993duration}. Like retrospective biases, prospective duration can also be biased: \citet{zauberman2009discounting} finds that subjective time perceptions are logarithmic in objective time \citep{zauberman2009discounting}, and \citet{liu1994mental} finds that time estimation was sensitive to the presence of cognitive demands. Various factors from existing research on subjective time perception could explain the speedup illusion. For example, previous studies find that conditions that favor vivid representations and efficient perceptual decision-making lead to longer perceived durations --- it is possible that people lack a vivid representation of task completions with AI and thus underestimate the time it would take \citep{matthews2016temporal}. Another important factor could be that people are conflating time reduction with effort reduction. A task can take a long time to complete --- even with the help of AI --- but the process could feel effortless. All the user had to do was write out the prompt and wait for the model to come up with an answer. This very difference could explain why a speedup illusion persists: when there is AI help, a task just does not \textit{feel} long, but when the help is gone, the task \textit{feels} long to complete. It is likely that when people offload, they start to miscalibrate about other resources being conserved, such as time. Our findings highlight the subjective nature of effort and how time reduction alone cannot fully capture the experienced efficiency gain. Besides time reduction, narratives and reports about AI productivity should also consider a more encompassing notion of cognitive offloading that takes into account the subjective experience.

\paragraph{A feedback loop: risks of overreliance.}

In this paper, we identify a speedup illusion of AI, which persists across task domains and difficulty levels, meaning that people expect AI to make task completion faster even when it does not. Studying calibration in the context of AI use is particularly important because it sheds lights on the factors that shape people's decision to (and potentially continue to) use AI systems. As resource-constrained reasoners~\citep{griffiths2020understanding}, if people expect AI assistance to reduce time, they are more likely to delegate tasks to AI. As cautioned in recent studies, an excessive amount of offloading could lead to cognitive deskilling \citep{oktar2025identifying, gerlich2025ai, ahn2025preserving}, which will impair people's cognitive capabilities and further slow down their independent completion time. The tendency risks forming a negative feedback loop where people use LLMs as steroids and offload at a disruptive level \citep{hofman2023steroids, jose2025outsourcing, collins2025revisiting}, which could undermine people's own cognitive abilities and foster overreliance~\citep{ibrahim2025measuring}. Even more jarringly, we provide preliminary evidence for a feedback loop where miscalibration encourages AI use as people overestimate the benefits of AI assistance, and using AI and experiencing effort saving leads to further miscalibration, which will lead to increased AI use.

While blindly copying prompts and pasting model responses significantly reduced cognitive effort, it could provide a shortcut that ultimately leads to cognitive surrender and disempowerment \citep{shaw2026thinking, sturgeon2025humanagencybench}. Our results on the distinction between time and effort reduction also show that people need to re-evaluate what resource (time versus effort) that they are trying to conserve. Yet, our qualitative analysis also indicates that many participants prompted LLMs in ways that required them to think, pointing to the possibility of LLMs allowing users to critically engage with the questions \citep{shum2024generative}. Similar to how LLMs can create an illusion of understanding, our work highlights people's skewed mental models and invite future work to more systematically examine the downstream outcomes of such misalignments \citep{messeri2024artificial, bansal2019beyond, kelly2023capturing}. More importantly, our work highlights the importance of calibration as both a subject and mechanism of study for understanding AI use and human-AI interaction more broadly; calibration can also be a key intervention point to adjust behavioral outcomes such as overreliance.

\subsection{Limitations and Future Directions}

Our study has various limitations that point to opportunities for future work. First, we focus on tasks that can be completed under 5 minutes; an interesting direction of future work is to identify the specific complexity level at which AI assistance genuinely saves time and effort. Second, our experiment did not directly control for participant motivation or incentives. Although we manually went through the responses and excluded low-effort and incorrect responses, crowdworking contexts may impact  task completion time, response quality, and AI adoption rates compared to other settings with more personal stakes \citep{yurt2024questionnaire}. 

Nonetheless, this setting enables us to understand overall miscalibrations in AI use. Future work should examine how incentive structures and motivation levels (e.g. answer quality) interact with the completion times and reported effort. Third, another limitation of the study is individual variations in AI use. Participants used AI in very different ways, and the degree of cognitive effort required even within the AI condition could vary. For example, it is possible that participants prompted the model but came up with a different final answer, regardless of what the AI response was. It was also possible that some participants only used AI to double check their answer or prompted AI out of curiosity of the response without using the responses. 

In our completion sample, we assign participants to one of the two completion modes. Future work should investigate \textit{when} people choose to use AI to understand the rate at which people are using AI and what resource (e.g. effort vs. time) they preserve, as well as the underlying mechanisms that lead to the decision to or not to use AI \citep{collins2024modulating}.

Finally, our study relies on between-subjects aggregates to avoid carryover biases. The prediction sample and completion sample were also not completely disjoint, with some participants completing both samples on Prolific. Future work should examine within-subject patterns while minimizing biases in the dependent variables.

\subsection{Conclusion}
Through a large-scale behavioral study on cognitive offloading, we highlight an asymmetric miscalibration and show that AI assistance can create a speedup illusion where tasks feel shorter to complete because the process becomes less effortful. We point out the dissociation between time and effort, finding that AI assistance always leads to effort reduction but not time reduction.

\clearpage

\section{Acknowledgments}

We would like to thank members of the Social Interaction Lab (SoIL) and the Jurafsky lab for their valuable feedback on the project. In addition, we thank Polina Tsvilodub and Charley Wu for their helpful feedback on the draft.

\bibliographystyle{bpacite}

\setlength{\bibleftmargin}{.125in}
\setlength{\bibindent}{-\bibleftmargin}

\bibliography{references.bib}

\begin{thebibliography}{}

\bibitem [\protect \citeauthoryear {%
Ahn%
}{%
Ahn%
}{%
{\protect \APACyear {2025}}%
}]{%
ahn2025preserving}
\APACinsertmetastar {%
ahn2025preserving}%
\begin{APACrefauthors}%
Ahn, S.%
\end{APACrefauthors}%
\unskip\
\newblock
\APACrefYearMonthDay{2025}{}{}.
\newblock
{\BBOQ}\APACrefatitle {Preserving Critical Thinking in the Age of Large Language Models: The Paradox of Cognitive Load and Efficiency} {Preserving critical thinking in the age of large language models: The paradox of cognitive load and efficiency}.{\BBCQ}
\newblock
\APACjournalVolNumPages{The Korean Journal of Medicine}{100}{5}{197--200}.
\PrintBackRefs{\CurrentBib}

\bibitem [\protect \citeauthoryear {%
Appel%
\ \protect \BOthers {.}}{%
Appel%
\ \protect \BOthers {.}}{%
{\protect \APACyear {2026}}%
}]{%
anthropic2026aeiv4}
\APACinsertmetastar {%
anthropic2026aeiv4}%
\begin{APACrefauthors}%
Appel, R.%
, Massenkoff, M.%
, McCrory, P.%
, McCain, M.%
, Heller, R.%
, Neylon, T.%
\BCBL {}\ \BBA {} Tamkin, A.%
\end{APACrefauthors}%
\unskip\
\newblock
\APACrefYearMonthDay{2026}{}{}.
\newblock
\APACrefbtitle {Anthropic Economic Index report: economic primitives.} {Anthropic economic index report: economic primitives.}
\newblock
\begin{APACrefURL} \url{https://www.anthropic.com/research/anthropic-economic-index-january-2026-report} \end{APACrefURL}
\PrintBackRefs{\CurrentBib}

\bibitem [\protect \citeauthoryear {%
Bansal%
\ \protect \BOthers {.}}{%
Bansal%
\ \protect \BOthers {.}}{%
{\protect \APACyear {2019}}%
}]{%
bansal2019beyond}
\APACinsertmetastar {%
bansal2019beyond}%
\begin{APACrefauthors}%
Bansal, G.%
, Nushi, B.%
, Kamar, E.%
, Lasecki, W\BPBI S.%
, Weld, D\BPBI S.%
\BCBL {}\ \BBA {} Horvitz, E.%
\end{APACrefauthors}%
\unskip\
\newblock
\APACrefYearMonthDay{2019}{}{}.
\newblock
{\BBOQ}\APACrefatitle {Beyond accuracy: The role of mental models in human-AI team performance} {Beyond accuracy: The role of mental models in human-ai team performance}.{\BBCQ}
\newblock
\BIn{} \APACrefbtitle {Proceedings of the {AAAI} conference on human computation and crowdsourcing} {Proceedings of the {AAAI} conference on human computation and crowdsourcing}\ (\BVOL~7, \BPGS\ 2--11).
\PrintBackRefs{\CurrentBib}

\bibitem [\protect \citeauthoryear {%
Becker%
, Rush%
, Barnes%
\BCBL {}\ \BBA {} Rein%
}{%
Becker%
\ \protect \BOthers {.}}{%
{\protect \APACyear {2025}}%
}]{%
becker2025measuring}
\APACinsertmetastar {%
becker2025measuring}%
\begin{APACrefauthors}%
Becker, J.%
, Rush, N.%
, Barnes, E.%
\BCBL {}\ \BBA {} Rein, D.%
\end{APACrefauthors}%
\unskip\
\newblock
\APACrefYearMonthDay{2025}{07}{}.
\newblock
\APACrefbtitle {Measuring the Impact of Early-2025 AI on Experienced Open-Source Developer Productivity.} {Measuring the impact of early-2025 ai on experienced open-source developer productivity.}
\newblock
\APAChowpublished {\url{https://metr.org/blog/2025-07-10-early-2025-ai-experienced-os-dev-study/}}.
\PrintBackRefs{\CurrentBib}

\bibitem [\protect \citeauthoryear {%
Cacioppo%
\ \BBA {} Petty%
}{%
Cacioppo%
\ \BBA {} Petty%
}{%
{\protect \APACyear {1982}}%
}]{%
cacioppo1982need}
\APACinsertmetastar {%
cacioppo1982need}%
\begin{APACrefauthors}%
Cacioppo, J\BPBI T.%
\BCBT {}\ \BBA {} Petty, R\BPBI E.%
\end{APACrefauthors}%
\unskip\
\newblock
\APACrefYearMonthDay{1982}{}{}.
\newblock
{\BBOQ}\APACrefatitle {The need for cognition.} {The need for cognition.}{\BBCQ}
\newblock
\APACjournalVolNumPages{Journal of personality and social psychology}{42}{1}{116}.
\PrintBackRefs{\CurrentBib}

\bibitem [\protect \citeauthoryear {%
Collins%
, Bhatt%
\BCBL {}\ \BBA {} Sucholutsky%
}{%
Collins%
\ \protect \BOthers {.}}{%
{\protect \APACyear {2025}}%
}]{%
collins2025revisiting}
\APACinsertmetastar {%
collins2025revisiting}%
\begin{APACrefauthors}%
Collins, K\BPBI M.%
, Bhatt, U.%
\BCBL {}\ \BBA {} Sucholutsky, I.%
\end{APACrefauthors}%
\unskip\
\newblock
\APACrefYearMonthDay{2025}{}{}.
\newblock
{\BBOQ}\APACrefatitle {Revisiting Rogers' Paradox in the Context of Human-AI Interaction} {Revisiting rogers' paradox in the context of human-ai interaction}.{\BBCQ}
\newblock
\APACjournalVolNumPages{arXiv preprint arXiv:2501.10476}{}{}{}.
\PrintBackRefs{\CurrentBib}

\bibitem [\protect \citeauthoryear {%
Collins%
, Chen%
\BCBL {}\ \protect \BOthers {.}}{%
Collins%
, Chen%
\BCBL {}\ \protect \BOthers {.}}{%
{\protect \APACyear {2024}}%
}]{%
collins2024modulating}
\APACinsertmetastar {%
collins2024modulating}%
\begin{APACrefauthors}%
Collins, K\BPBI M.%
, Chen, V.%
, Sucholutsky, I.%
, Kirk, H\BPBI R.%
, Sadek, M.%
, Sargeant, H.%
\BDBL {}Bhatt, U.%
\end{APACrefauthors}%
\unskip\
\newblock
\APACrefYearMonthDay{2024}{}{}.
\newblock
{\BBOQ}\APACrefatitle {Modulating language model experiences through frictions} {Modulating language model experiences through frictions}.{\BBCQ}
\newblock
\APACjournalVolNumPages{arXiv preprint arXiv:2407.12804}{}{}{}.
\PrintBackRefs{\CurrentBib}

\bibitem [\protect \citeauthoryear {%
Collins%
, Sucholutsky%
\BCBL {}\ \protect \BOthers {.}}{%
Collins%
, Sucholutsky%
\BCBL {}\ \protect \BOthers {.}}{%
{\protect \APACyear {2024}}%
}]{%
collins2024building}
\APACinsertmetastar {%
collins2024building}%
\begin{APACrefauthors}%
Collins, K\BPBI M.%
, Sucholutsky, I.%
, Bhatt, U.%
, Chandra, K.%
, Wong, L.%
, Lee, M.%
\BDBL {}Griffiths, T\BPBI L.%
\end{APACrefauthors}%
\unskip\
\newblock
\APACrefYearMonthDay{2024}{}{}.
\newblock
{\BBOQ}\APACrefatitle {Building machines that learn and think with people} {Building machines that learn and think with people}.{\BBCQ}
\newblock
\APACjournalVolNumPages{Nature human behaviour}{8}{10}{1851--1863}.
\PrintBackRefs{\CurrentBib}

\bibitem [\protect \citeauthoryear {%
Dhillon%
\ \protect \BOthers {.}}{%
Dhillon%
\ \protect \BOthers {.}}{%
{\protect \APACyear {2024}}%
}]{%
dhillon2024shaping}
\APACinsertmetastar {%
dhillon2024shaping}%
\begin{APACrefauthors}%
Dhillon, P\BPBI S.%
, Molaei, S.%
, Li, J.%
, Golub, M.%
, Zheng, S.%
\BCBL {}\ \BBA {} Robert, L\BPBI P.%
\end{APACrefauthors}%
\unskip\
\newblock
\APACrefYearMonthDay{2024}{}{}.
\newblock
{\BBOQ}\APACrefatitle {Shaping human-{AI} collaboration: Varied scaffolding levels in co-writing with language models} {Shaping human-{AI} collaboration: Varied scaffolding levels in co-writing with language models}.{\BBCQ}
\newblock
\BIn{} \APACrefbtitle {Proceedings of the 2024 {CHI} Conference on Human Factors in Computing Systems} {Proceedings of the 2024 {CHI} conference on human factors in computing systems}\ (\BPGS\ 1--18).
\PrintBackRefs{\CurrentBib}

\bibitem [\protect \citeauthoryear {%
Dunn%
\ \BBA {} Risko%
}{%
Dunn%
\ \BBA {} Risko%
}{%
{\protect \APACyear {2016}}%
}]{%
dunn2016toward}
\APACinsertmetastar {%
dunn2016toward}%
\begin{APACrefauthors}%
Dunn, T\BPBI L.%
\BCBT {}\ \BBA {} Risko, E\BPBI F.%
\end{APACrefauthors}%
\unskip\
\newblock
\APACrefYearMonthDay{2016}{}{}.
\newblock
{\BBOQ}\APACrefatitle {Toward a metacognitive account of cognitive offloading} {Toward a metacognitive account of cognitive offloading}.{\BBCQ}
\newblock
\APACjournalVolNumPages{Cognitive Science}{40}{5}{1080--1127}.
\PrintBackRefs{\CurrentBib}

\bibitem [\protect \citeauthoryear {%
Fan%
, Bainbridge%
, Chamberlain%
\BCBL {}\ \BBA {} Wammes%
}{%
Fan%
\ \protect \BOthers {.}}{%
{\protect \APACyear {2023}}%
}]{%
fan2023drawing}
\APACinsertmetastar {%
fan2023drawing}%
\begin{APACrefauthors}%
Fan, J\BPBI E.%
, Bainbridge, W\BPBI A.%
, Chamberlain, R.%
\BCBL {}\ \BBA {} Wammes, J\BPBI D.%
\end{APACrefauthors}%
\unskip\
\newblock
\APACrefYearMonthDay{2023}{}{}.
\newblock
{\BBOQ}\APACrefatitle {Drawing as a versatile cognitive tool} {Drawing as a versatile cognitive tool}.{\BBCQ}
\newblock
\APACjournalVolNumPages{Nature Reviews Psychology}{2}{9}{556--568}.
\PrintBackRefs{\CurrentBib}

\bibitem [\protect \citeauthoryear {%
Fraisse%
\ \protect \BOthers {.}}{%
Fraisse%
\ \protect \BOthers {.}}{%
{\protect \APACyear {1984}}%
}]{%
fraisse1984perception}
\APACinsertmetastar {%
fraisse1984perception}%
\begin{APACrefauthors}%
Fraisse, P.%
\BCBT {}\ \BOthersPeriod {.}
\end{APACrefauthors}%
\unskip\
\newblock
\APACrefYearMonthDay{1984}{}{}.
\newblock
{\BBOQ}\APACrefatitle {Perception and estimation of time} {Perception and estimation of time}.{\BBCQ}
\newblock
\APACjournalVolNumPages{Annual review of psychology}{35}{1}{1--37}.
\PrintBackRefs{\CurrentBib}

\bibitem [\protect \citeauthoryear {%
Fredrickson%
\ \BBA {} Kahneman%
}{%
Fredrickson%
\ \BBA {} Kahneman%
}{%
{\protect \APACyear {1993}}%
}]{%
fredrickson1993duration}
\APACinsertmetastar {%
fredrickson1993duration}%
\begin{APACrefauthors}%
Fredrickson, B\BPBI L.%
\BCBT {}\ \BBA {} Kahneman, D.%
\end{APACrefauthors}%
\unskip\
\newblock
\APACrefYearMonthDay{1993}{}{}.
\newblock
{\BBOQ}\APACrefatitle {Duration neglect in retrospective evaluations of affective episodes.} {Duration neglect in retrospective evaluations of affective episodes.}{\BBCQ}
\newblock
\APACjournalVolNumPages{Journal of personality and social psychology}{65}{1}{45}.
\PrintBackRefs{\CurrentBib}

\bibitem [\protect \citeauthoryear {%
Gerlich%
}{%
Gerlich%
}{%
{\protect \APACyear {2025}}%
}]{%
gerlich2025ai}
\APACinsertmetastar {%
gerlich2025ai}%
\begin{APACrefauthors}%
Gerlich, M.%
\end{APACrefauthors}%
\unskip\
\newblock
\APACrefYearMonthDay{2025}{}{}.
\newblock
{\BBOQ}\APACrefatitle {{AI} tools in society: Impacts on cognitive offloading and the future of critical thinking} {{AI} tools in society: Impacts on cognitive offloading and the future of critical thinking}.{\BBCQ}
\newblock
\APACjournalVolNumPages{Societies}{15}{1}{6}.
\PrintBackRefs{\CurrentBib}

\bibitem [\protect \citeauthoryear {%
Grassini%
}{%
Grassini%
}{%
{\protect \APACyear {2023}}%
}]{%
grassini2023development}
\APACinsertmetastar {%
grassini2023development}%
\begin{APACrefauthors}%
Grassini, S.%
\end{APACrefauthors}%
\unskip\
\newblock
\APACrefYearMonthDay{2023}{}{}.
\newblock
{\BBOQ}\APACrefatitle {Development and validation of the {AI} attitude scale (AIAS-4): a brief measure of general attitude toward artificial intelligence} {Development and validation of the {AI} attitude scale (aias-4): a brief measure of general attitude toward artificial intelligence}.{\BBCQ}
\newblock
\APACjournalVolNumPages{Frontiers in psychology}{14}{}{1191628}.
\PrintBackRefs{\CurrentBib}

\bibitem [\protect \citeauthoryear {%
Griffiths%
}{%
Griffiths%
}{%
{\protect \APACyear {2020}}%
}]{%
griffiths2020understanding}
\APACinsertmetastar {%
griffiths2020understanding}%
\begin{APACrefauthors}%
Griffiths, T\BPBI L.%
\end{APACrefauthors}%
\unskip\
\newblock
\APACrefYearMonthDay{2020}{}{}.
\newblock
{\BBOQ}\APACrefatitle {Understanding human intelligence through human limitations} {Understanding human intelligence through human limitations}.{\BBCQ}
\newblock
\APACjournalVolNumPages{Trends in Cognitive Sciences}{24}{11}{873--883}.
\PrintBackRefs{\CurrentBib}

\bibitem [\protect \citeauthoryear {%
Griffiths%
\ \protect \BOthers {.}}{%
Griffiths%
\ \protect \BOthers {.}}{%
{\protect \APACyear {2019}}%
}]{%
griffiths2019doing}
\APACinsertmetastar {%
griffiths2019doing}%
\begin{APACrefauthors}%
Griffiths, T\BPBI L.%
, Callaway, F.%
, Chang, M\BPBI B.%
, Grant, E.%
, Krueger, P\BPBI M.%
\BCBL {}\ \BBA {} Lieder, F.%
\end{APACrefauthors}%
\unskip\
\newblock
\APACrefYearMonthDay{2019}{}{}.
\newblock
{\BBOQ}\APACrefatitle {Doing more with less: meta-reasoning and meta-learning in humans and machines} {Doing more with less: meta-reasoning and meta-learning in humans and machines}.{\BBCQ}
\newblock
\APACjournalVolNumPages{Current Opinion in Behavioral Sciences}{29}{}{24--30}.
\PrintBackRefs{\CurrentBib}

\bibitem [\protect \citeauthoryear {%
Handa%
\ \protect \BOthers {.}}{%
Handa%
\ \protect \BOthers {.}}{%
{\protect \APACyear {2025}}%
}]{%
handa2025economic}
\APACinsertmetastar {%
handa2025economic}%
\begin{APACrefauthors}%
Handa, K.%
, Tamkin, A.%
, McCain, M.%
, Huang, S.%
, Durmus, E.%
, Heck, S.%
\BDBL {}others%
\end{APACrefauthors}%
\unskip\
\newblock
\APACrefYearMonthDay{2025}{}{}.
\newblock
{\BBOQ}\APACrefatitle {Which economic tasks are performed with ai? evidence from millions of claude conversations} {Which economic tasks are performed with ai? evidence from millions of claude conversations}.{\BBCQ}
\newblock
\APACjournalVolNumPages{arXiv preprint arXiv:2503.04761}{}{}{}.
\PrintBackRefs{\CurrentBib}

\bibitem [\protect \citeauthoryear {%
Hart%
\ \BBA {} Staveland%
}{%
Hart%
\ \BBA {} Staveland%
}{%
{\protect \APACyear {1988}}%
}]{%
Hart_1986}
\APACinsertmetastar {%
Hart_1986}%
\begin{APACrefauthors}%
Hart, S\BPBI G.%
\BCBT {}\ \BBA {} Staveland, L\BPBI E.%
\end{APACrefauthors}%
\unskip\
\newblock
\APACrefYearMonthDay{1988}{}{}.
\newblock
{\BBOQ}\APACrefatitle {Development of NASA-TLX (Task Load Index): Results of empirical and theoretical research} {Development of nasa-tlx (task load index): Results of empirical and theoretical research}.{\BBCQ}
\newblock
\BIn{} \APACrefbtitle {Advances in psychology} {Advances in psychology}\ (\BVOL~52, \BPGS\ 139--183).
\newblock
\APACaddressPublisher{}{Elsevier}.
\PrintBackRefs{\CurrentBib}

\bibitem [\protect \citeauthoryear {%
Hofman%
, Goldstein%
\BCBL {}\ \BBA {} Rothschild%
}{%
Hofman%
\ \protect \BOthers {.}}{%
{\protect \APACyear {2023}}%
}]{%
hofman2023steroids}
\APACinsertmetastar {%
hofman2023steroids}%
\begin{APACrefauthors}%
Hofman, J.%
, Goldstein, D\BPBI G.%
\BCBL {}\ \BBA {} Rothschild, D.%
\end{APACrefauthors}%
\unskip\
\newblock
\APACrefYearMonthDay{2023}{December}{}.
\newblock
{\BBOQ}\APACrefatitle {A Sports Analogy for Understanding Different Ways to Use AI} {A sports analogy for understanding different ways to use ai}.{\BBCQ}
\newblock
\APACjournalVolNumPages{Harvard Business Review}{}{}{}.
\newblock
\begin{APACrefURL} \url{https://www.microsoft.com/en-us/research/publication/a-sports-analogy-for-understanding-different-ways-to-use-ai/} \end{APACrefURL}
\PrintBackRefs{\CurrentBib}

\bibitem [\protect \citeauthoryear {%
Hooper%
}{%
Hooper%
}{%
{\protect \APACyear {2025}}%
}]{%
hooper2025cognitive}
\APACinsertmetastar {%
hooper2025cognitive}%
\begin{APACrefauthors}%
Hooper, V\BPBI J.%
\end{APACrefauthors}%
\unskip\
\newblock
\APACrefYearMonthDay{2025}{}{}.
\newblock
{\BBOQ}\APACrefatitle {Cognitive offloading and the reshaping of human thought: The subtle influence of Artificial Intelligence} {Cognitive offloading and the reshaping of human thought: The subtle influence of artificial intelligence}.{\BBCQ}
\newblock
\BIn{} \APACrefbtitle {Colloquia, Academic Journal of Culture and Thought} {Colloquia, academic journal of culture and thought}\ (\BVOL~12, \BPGS\ 01--14).
\PrintBackRefs{\CurrentBib}

\bibitem [\protect \citeauthoryear {%
Ibrahim%
\ \protect \BOthers {.}}{%
Ibrahim%
\ \protect \BOthers {.}}{%
{\protect \APACyear {2025}}%
}]{%
ibrahim2025measuring}
\APACinsertmetastar {%
ibrahim2025measuring}%
\begin{APACrefauthors}%
Ibrahim, L.%
, Collins, K\BPBI M.%
, Kim, S\BPBI S\BPBI Y.%
, Reuel, A.%
, Lamparth, M.%
, Feng, K.%
\BDBL {}Bhatt, U.%
\end{APACrefauthors}%
\unskip\
\newblock
\APACrefYearMonthDay{2025}{}{}.
\newblock
{\BBOQ}\APACrefatitle {Measuring and mitigating overreliance is necessary for building human-compatible {AI}} {Measuring and mitigating overreliance is necessary for building human-compatible {AI}}.{\BBCQ}
\newblock
\APACjournalVolNumPages{arXiv preprint arXiv:2509.08010}{}{}{}.
\PrintBackRefs{\CurrentBib}

\bibitem [\protect \citeauthoryear {%
Jose%
\ \protect \BOthers {.}}{%
Jose%
\ \protect \BOthers {.}}{%
{\protect \APACyear {2025}}%
}]{%
jose2025outsourcing}
\APACinsertmetastar {%
jose2025outsourcing}%
\begin{APACrefauthors}%
Jose, B.%
, Joseph, D.%
, Mohan, V.%
, Alexander, E.%
, Varghese, S\BPBI K.%
\BCBL {}\ \BBA {} Roy, A.%
\end{APACrefauthors}%
\unskip\
\newblock
\APACrefYearMonthDay{2025}{}{}.
\newblock
{\BBOQ}\APACrefatitle {Outsourcing cognition: the psychological costs of AI-era convenience} {Outsourcing cognition: the psychological costs of ai-era convenience}.{\BBCQ}
\newblock
\APACjournalVolNumPages{Frontiers in Psychology}{16}{}{1645237}.
\PrintBackRefs{\CurrentBib}

\bibitem [\protect \citeauthoryear {%
Kelly%
, Kumar%
, Smyth%
\BCBL {}\ \BBA {} Steyvers%
}{%
Kelly%
\ \protect \BOthers {.}}{%
{\protect \APACyear {2023}}%
}]{%
kelly2023capturing}
\APACinsertmetastar {%
kelly2023capturing}%
\begin{APACrefauthors}%
Kelly, M.%
, Kumar, A.%
, Smyth, P.%
\BCBL {}\ \BBA {} Steyvers, M.%
\end{APACrefauthors}%
\unskip\
\newblock
\APACrefYearMonthDay{2023}{}{}.
\newblock
{\BBOQ}\APACrefatitle {Capturing Humans’ mental models of {AI}: An item response theory approach} {Capturing humans’ mental models of {AI}: An item response theory approach}.{\BBCQ}
\newblock
\BIn{} \APACrefbtitle {Proceedings of the 2023 {ACM} conference on fairness, accountability, and transparency} {Proceedings of the 2023 {ACM} conference on fairness, accountability, and transparency}\ (\BPGS\ 1723--1734).
\PrintBackRefs{\CurrentBib}

\bibitem [\protect \citeauthoryear {%
Koriat%
}{%
Koriat%
}{%
{\protect \APACyear {2015}}%
}]{%
koriat2015metacognition}
\APACinsertmetastar {%
koriat2015metacognition}%
\begin{APACrefauthors}%
Koriat, A.%
\end{APACrefauthors}%
\unskip\
\newblock
\APACrefYearMonthDay{2015}{}{}.
\newblock
{\BBOQ}\APACrefatitle {Metacognition: Decision making Processes in Self-monitoring and Self-regulation} {Metacognition: Decision making processes in self-monitoring and self-regulation}.{\BBCQ}
\newblock
\APACjournalVolNumPages{The Wiley Blackwell handbook of judgment and decision making}{2}{}{356--379}.
\PrintBackRefs{\CurrentBib}

\bibitem [\protect \citeauthoryear {%
Korzynski%
, Mazurek%
, Krzypkowska%
\BCBL {}\ \BBA {} Kurasinski%
}{%
Korzynski%
\ \protect \BOthers {.}}{%
{\protect \APACyear {2023}}%
}]{%
korzynski2023artificial}
\APACinsertmetastar {%
korzynski2023artificial}%
\begin{APACrefauthors}%
Korzynski, P.%
, Mazurek, G.%
, Krzypkowska, P.%
\BCBL {}\ \BBA {} Kurasinski, A.%
\end{APACrefauthors}%
\unskip\
\newblock
\APACrefYearMonthDay{2023}{}{}.
\newblock
{\BBOQ}\APACrefatitle {Artificial intelligence prompt engineering as a new digital competence: Analysis of generative {AI} technologies such as ChatGPT} {Artificial intelligence prompt engineering as a new digital competence: Analysis of generative {AI} technologies such as chatgpt}.{\BBCQ}
\newblock
\APACjournalVolNumPages{Entrepreneurial Business and Economics Review}{11}{3}{25--37}.
\PrintBackRefs{\CurrentBib}

\bibitem [\protect \citeauthoryear {%
Lieder%
\ \BBA {} Griffiths%
}{%
Lieder%
\ \BBA {} Griffiths%
}{%
{\protect \APACyear {2020}}%
}]{%
lieder2020resource}
\APACinsertmetastar {%
lieder2020resource}%
\begin{APACrefauthors}%
Lieder, F.%
\BCBT {}\ \BBA {} Griffiths, T\BPBI L.%
\end{APACrefauthors}%
\unskip\
\newblock
\APACrefYearMonthDay{2020}{}{}.
\newblock
{\BBOQ}\APACrefatitle {Resource-rational analysis: Understanding human cognition as the optimal use of limited computational resources} {Resource-rational analysis: Understanding human cognition as the optimal use of limited computational resources}.{\BBCQ}
\newblock
\APACjournalVolNumPages{Behavioral and brain sciences}{43}{}{e1}.
\PrintBackRefs{\CurrentBib}

\bibitem [\protect \citeauthoryear {%
Liu%
\ \BBA {} Wickens%
}{%
Liu%
\ \BBA {} Wickens%
}{%
{\protect \APACyear {1994}}%
}]{%
liu1994mental}
\APACinsertmetastar {%
liu1994mental}%
\begin{APACrefauthors}%
Liu, Y.%
\BCBT {}\ \BBA {} Wickens, C\BPBI D.%
\end{APACrefauthors}%
\unskip\
\newblock
\APACrefYearMonthDay{1994}{}{}.
\newblock
{\BBOQ}\APACrefatitle {Mental workload and cognitive task automaticity: an evaluation of subjective and time estimation metrics} {Mental workload and cognitive task automaticity: an evaluation of subjective and time estimation metrics}.{\BBCQ}
\newblock
\APACjournalVolNumPages{Ergonomics}{37}{11}{1843--1854}.
\PrintBackRefs{\CurrentBib}

\bibitem [\protect \citeauthoryear {%
Matthews%
\ \BBA {} Meck%
}{%
Matthews%
\ \BBA {} Meck%
}{%
{\protect \APACyear {2016}}%
}]{%
matthews2016temporal}
\APACinsertmetastar {%
matthews2016temporal}%
\begin{APACrefauthors}%
Matthews, W\BPBI J.%
\BCBT {}\ \BBA {} Meck, W\BPBI H.%
\end{APACrefauthors}%
\unskip\
\newblock
\APACrefYearMonthDay{2016}{}{}.
\newblock
{\BBOQ}\APACrefatitle {Temporal cognition: Connecting subjective time to perception, attention, and memory.} {Temporal cognition: Connecting subjective time to perception, attention, and memory.}{\BBCQ}
\newblock
\APACjournalVolNumPages{Psychological bulletin}{142}{8}{865}.
\PrintBackRefs{\CurrentBib}

\bibitem [\protect \citeauthoryear {%
Messeri%
\ \BBA {} Crockett%
}{%
Messeri%
\ \BBA {} Crockett%
}{%
{\protect \APACyear {2024}}%
}]{%
messeri2024artificial}
\APACinsertmetastar {%
messeri2024artificial}%
\begin{APACrefauthors}%
Messeri, L.%
\BCBT {}\ \BBA {} Crockett, M\BPBI J.%
\end{APACrefauthors}%
\unskip\
\newblock
\APACrefYearMonthDay{2024}{}{}.
\newblock
{\BBOQ}\APACrefatitle {Artificial intelligence and illusions of understanding in scientific research} {Artificial intelligence and illusions of understanding in scientific research}.{\BBCQ}
\newblock
\APACjournalVolNumPages{Nature}{627}{8002}{49--58}.
\PrintBackRefs{\CurrentBib}

\bibitem [\protect \citeauthoryear {%
Oktar%
\ \protect \BOthers {.}}{%
Oktar%
\ \protect \BOthers {.}}{%
{\protect \APACyear {2026}}%
}]{%
oktar2025identifying}
\APACinsertmetastar {%
oktar2025identifying}%
\begin{APACrefauthors}%
Oktar, K.%
, Collins, K\BPBI M.%
, Hern\'{a}ndez-Orallo, J.%
, Coyle, D.%
, Cave, S.%
, Weller, A.%
\BCBL {}\ \BBA {} Sucholutsky, I.%
\end{APACrefauthors}%
\unskip\
\newblock
\APACrefYearMonthDay{2026}{{\APACmonth{03}}}{}.
\newblock
{\BBOQ}\APACrefatitle {Identifying, Evaluating, and Mitigating Risks of AI Thought Partnerships} {Identifying, evaluating, and mitigating risks of ai thought partnerships}.{\BBCQ}
\newblock
\APACjournalVolNumPages{ACM AI Lett.}{}{}{}.
\newblock
\begin{APACrefURL} \url{https://doi.org/10.1145/3803024} \end{APACrefURL}
\newblock
\begin{APACrefDOI} \doi{10.1145/3803024} \end{APACrefDOI}
\PrintBackRefs{\CurrentBib}

\bibitem [\protect \citeauthoryear {%
Overgaard%
\ \BBA {} Sandberg%
}{%
Overgaard%
\ \BBA {} Sandberg%
}{%
{\protect \APACyear {2012}}%
}]{%
overgaard2012kinds}
\APACinsertmetastar {%
overgaard2012kinds}%
\begin{APACrefauthors}%
Overgaard, M.%
\BCBT {}\ \BBA {} Sandberg, K.%
\end{APACrefauthors}%
\unskip\
\newblock
\APACrefYearMonthDay{2012}{}{}.
\newblock
{\BBOQ}\APACrefatitle {Kinds of access: different methods for report reveal different kinds of metacognitive access} {Kinds of access: different methods for report reveal different kinds of metacognitive access}.{\BBCQ}
\newblock
\APACjournalVolNumPages{Philosophical Transactions of the Royal Society B: Biological Sciences}{367}{1594}{1287--1296}.
\PrintBackRefs{\CurrentBib}

\bibitem [\protect \citeauthoryear {%
Risko%
\ \BBA {} Gilbert%
}{%
Risko%
\ \BBA {} Gilbert%
}{%
{\protect \APACyear {2016}}%
}]{%
risko2016cognitive}
\APACinsertmetastar {%
risko2016cognitive}%
\begin{APACrefauthors}%
Risko, E\BPBI F.%
\BCBT {}\ \BBA {} Gilbert, S\BPBI J.%
\end{APACrefauthors}%
\unskip\
\newblock
\APACrefYearMonthDay{2016}{}{}.
\newblock
{\BBOQ}\APACrefatitle {Cognitive offloading} {Cognitive offloading}.{\BBCQ}
\newblock
\APACjournalVolNumPages{Trends in cognitive sciences}{20}{9}{676--688}.
\PrintBackRefs{\CurrentBib}

\bibitem [\protect \citeauthoryear {%
Shaw%
\ \BBA {} Nave%
}{%
Shaw%
\ \BBA {} Nave%
}{%
{\protect \APACyear {2026}}%
}]{%
shaw2026thinking}
\APACinsertmetastar {%
shaw2026thinking}%
\begin{APACrefauthors}%
Shaw, S.%
\BCBT {}\ \BBA {} Nave, G.%
\end{APACrefauthors}%
\unskip\
\newblock
\APACrefYearMonthDay{2026}{}{}.
\newblock
{\BBOQ}\APACrefatitle {Thinking—Fast, slow, and artificial: How AI is reshaping human reasoning and the rise of cognitive surrender. PsyArXiv} {Thinking—fast, slow, and artificial: How ai is reshaping human reasoning and the rise of cognitive surrender. psyarxiv}.{\BBCQ}
\newblock
\APACjournalVolNumPages{Preprint at https://osf. io/preprints/psyarxiv/yk25n\_v1}{}{}{}.
\PrintBackRefs{\CurrentBib}

\bibitem [\protect \citeauthoryear {%
Shelby%
, Diaz%
\BCBL {}\ \BBA {} Prabhakaran%
}{%
Shelby%
\ \protect \BOthers {.}}{%
{\protect \APACyear {2025}}%
}]{%
shelby2025taxonomy}
\APACinsertmetastar {%
shelby2025taxonomy}%
\begin{APACrefauthors}%
Shelby, R.%
, Diaz, F.%
\BCBL {}\ \BBA {} Prabhakaran, V.%
\end{APACrefauthors}%
\unskip\
\newblock
\APACrefYearMonthDay{2025}{}{}.
\newblock
{\BBOQ}\APACrefatitle {Taxonomy of User Needs and Actions} {Taxonomy of user needs and actions}.{\BBCQ}
\newblock
\APACjournalVolNumPages{arXiv preprint arXiv:2510.06124}{}{}{}.
\PrintBackRefs{\CurrentBib}

\bibitem [\protect \citeauthoryear {%
Shum%
}{%
Shum%
}{%
{\protect \APACyear {2024}}%
}]{%
shum2024generative}
\APACinsertmetastar {%
shum2024generative}%
\begin{APACrefauthors}%
Shum, S\BPBI B.%
\end{APACrefauthors}%
\unskip\
\newblock
\APACrefYearMonthDay{2024}{}{}.
\newblock
{\BBOQ}\APACrefatitle {Generative AI for Critical Analysis Practical Tools Cognitive Offloading and Human Agency} {Generative ai for critical analysis practical tools cognitive offloading and human agency}.{\BBCQ}
\newblock
\BIn{} \APACrefbtitle {{CEUR} Workshop Proceedings.} {{CEUR} workshop proceedings.}
\PrintBackRefs{\CurrentBib}

\bibitem [\protect \citeauthoryear {%
Stadler%
, Bannert%
\BCBL {}\ \BBA {} Sailer%
}{%
Stadler%
\ \protect \BOthers {.}}{%
{\protect \APACyear {2024}}%
}]{%
stadler2024cognitive}
\APACinsertmetastar {%
stadler2024cognitive}%
\begin{APACrefauthors}%
Stadler, M.%
, Bannert, M.%
\BCBL {}\ \BBA {} Sailer, M.%
\end{APACrefauthors}%
\unskip\
\newblock
\APACrefYearMonthDay{2024}{}{}.
\newblock
{\BBOQ}\APACrefatitle {Cognitive ease at a cost: LLMs reduce mental effort but compromise depth in student scientific inquiry} {Cognitive ease at a cost: Llms reduce mental effort but compromise depth in student scientific inquiry}.{\BBCQ}
\newblock
\APACjournalVolNumPages{Computers in Human Behavior}{160}{}{108386}.
\PrintBackRefs{\CurrentBib}

\bibitem [\protect \citeauthoryear {%
Steele%
}{%
Steele%
}{%
{\protect \APACyear {2020}}%
}]{%
steele2020perception}
\APACinsertmetastar {%
steele2020perception}%
\begin{APACrefauthors}%
Steele, J.%
\end{APACrefauthors}%
\unskip\
\newblock
\APACrefYearMonthDay{2020}{Jun}{}.
\newblock
\APACrefbtitle {What is (perception of) effort? Objective and subjective effort during attempted task performance.} {What is (perception of) effort? objective and subjective effort during attempted task performance.}
\newblock
\APACaddressPublisher{}{PsyArXiv}.
\newblock
\begin{APACrefURL} \url{osf.io/preprints/psyarxiv/kbyhm_v1} \end{APACrefURL}
\newblock
\begin{APACrefDOI} \doi{10.31234/osf.io/kbyhm} \end{APACrefDOI}
\PrintBackRefs{\CurrentBib}

\bibitem [\protect \citeauthoryear {%
Sturgeon%
, Samuelson%
, Haimes%
\BCBL {}\ \BBA {} Anthis%
}{%
Sturgeon%
\ \protect \BOthers {.}}{%
{\protect \APACyear {2025}}%
}]{%
sturgeon2025humanagencybench}
\APACinsertmetastar {%
sturgeon2025humanagencybench}%
\begin{APACrefauthors}%
Sturgeon, B.%
, Samuelson, D.%
, Haimes, J.%
\BCBL {}\ \BBA {} Anthis, J\BPBI R.%
\end{APACrefauthors}%
\unskip\
\newblock
\APACrefYearMonthDay{2025}{}{}.
\newblock
{\BBOQ}\APACrefatitle {HumanAgencyBench: Scalable Evaluation of Human Agency Support in AI Assistants} {Humanagencybench: Scalable evaluation of human agency support in ai assistants}.{\BBCQ}
\newblock
\APACjournalVolNumPages{arXiv preprint arXiv:2509.08494}{}{}{}.
\PrintBackRefs{\CurrentBib}

\bibitem [\protect \citeauthoryear {%
Wahn%
, Schmitz%
, Gerster%
\BCBL {}\ \BBA {} Weiss%
}{%
Wahn%
\ \protect \BOthers {.}}{%
{\protect \APACyear {2023}}%
}]{%
wahn2023offloading}
\APACinsertmetastar {%
wahn2023offloading}%
\begin{APACrefauthors}%
Wahn, B.%
, Schmitz, L.%
, Gerster, F\BPBI N.%
\BCBL {}\ \BBA {} Weiss, M.%
\end{APACrefauthors}%
\unskip\
\newblock
\APACrefYearMonthDay{2023}{}{}.
\newblock
{\BBOQ}\APACrefatitle {Offloading under cognitive load: Humans are willing to offload parts of an attentionally demanding task to an algorithm} {Offloading under cognitive load: Humans are willing to offload parts of an attentionally demanding task to an algorithm}.{\BBCQ}
\newblock
\APACjournalVolNumPages{Plos one}{18}{5}{e0286102}.
\PrintBackRefs{\CurrentBib}

\bibitem [\protect \citeauthoryear {%
Wang%
\ \protect \BOthers {.}}{%
Wang%
\ \protect \BOthers {.}}{%
{\protect \APACyear {2025}}%
}]{%
wang2025ai}
\APACinsertmetastar {%
wang2025ai}%
\begin{APACrefauthors}%
Wang, Z\BPBI Z.%
, Shao, Y.%
, Shaikh, O.%
, Fried, D.%
, Neubig, G.%
\BCBL {}\ \BBA {} Yang, D.%
\end{APACrefauthors}%
\unskip\
\newblock
\APACrefYearMonthDay{2025}{}{}.
\newblock
{\BBOQ}\APACrefatitle {How Do {AI} Agents Do Human Work? Comparing AI and Human Workflows Across Diverse Occupations} {How do {AI} agents do human work? comparing ai and human workflows across diverse occupations}.{\BBCQ}
\newblock
\APACjournalVolNumPages{arXiv preprint arXiv:2510.22780}{}{}{}.
\PrintBackRefs{\CurrentBib}

\bibitem [\protect \citeauthoryear {%
Xiong%
\ \protect \BOthers {.}}{%
Xiong%
\ \protect \BOthers {.}}{%
{\protect \APACyear {2024}}%
}]{%
xiong2024search}
\APACinsertmetastar {%
xiong2024search}%
\begin{APACrefauthors}%
Xiong, H.%
, Bian, J.%
, Li, Y.%
, Li, X.%
, Du, M.%
, Wang, S.%
\BDBL {}Helal, S.%
\end{APACrefauthors}%
\unskip\
\newblock
\APACrefYearMonthDay{2024}{}{}.
\newblock
{\BBOQ}\APACrefatitle {When search engine services meet large language models: visions and challenges} {When search engine services meet large language models: visions and challenges}.{\BBCQ}
\newblock
\APACjournalVolNumPages{IEEE Transactions on Services Computing}{}{}{}.
\PrintBackRefs{\CurrentBib}

\bibitem [\protect \citeauthoryear {%
Yeo%
\ \BBA {} Neal%
}{%
Yeo%
\ \BBA {} Neal%
}{%
{\protect \APACyear {2008}}%
}]{%
yeo2008subjective}
\APACinsertmetastar {%
yeo2008subjective}%
\begin{APACrefauthors}%
Yeo, G.%
\BCBT {}\ \BBA {} Neal, A.%
\end{APACrefauthors}%
\unskip\
\newblock
\APACrefYearMonthDay{2008}{}{}.
\newblock
{\BBOQ}\APACrefatitle {Subjective cognitive effort: A model of states, traits, and time.} {Subjective cognitive effort: A model of states, traits, and time.}{\BBCQ}
\newblock
\APACjournalVolNumPages{Journal of Applied Psychology}{93}{3}{617}.
\PrintBackRefs{\CurrentBib}

\bibitem [\protect \citeauthoryear {%
Yurt%
\ \BBA {} Kasarci%
}{%
Yurt%
\ \BBA {} Kasarci%
}{%
{\protect \APACyear {2024}}%
}]{%
yurt2024questionnaire}
\APACinsertmetastar {%
yurt2024questionnaire}%
\begin{APACrefauthors}%
Yurt, E.%
\BCBT {}\ \BBA {} Kasarci, I.%
\end{APACrefauthors}%
\unskip\
\newblock
\APACrefYearMonthDay{2024}{}{}.
\newblock
{\BBOQ}\APACrefatitle {A Questionnaire of Artificial Intelligence Use Motives: A contribution to investigating the connection between AI and motivation} {A questionnaire of artificial intelligence use motives: A contribution to investigating the connection between ai and motivation}.{\BBCQ}
\newblock
\APACjournalVolNumPages{International Journal of Technology in Education}{7}{2}{}.
\PrintBackRefs{\CurrentBib}

\bibitem [\protect \citeauthoryear {%
Zauberman%
, Kim%
, Malkoc%
\BCBL {}\ \BBA {} Bettman%
}{%
Zauberman%
\ \protect \BOthers {.}}{%
{\protect \APACyear {2009}}%
}]{%
zauberman2009discounting}
\APACinsertmetastar {%
zauberman2009discounting}%
\begin{APACrefauthors}%
Zauberman, G.%
, Kim, B\BPBI K.%
, Malkoc, S\BPBI A.%
\BCBL {}\ \BBA {} Bettman, J\BPBI R.%
\end{APACrefauthors}%
\unskip\
\newblock
\APACrefYearMonthDay{2009}{}{}.
\newblock
{\BBOQ}\APACrefatitle {Discounting time and time discounting: Subjective time perception and intertemporal preferences} {Discounting time and time discounting: Subjective time perception and intertemporal preferences}.{\BBCQ}
\newblock
\APACjournalVolNumPages{Journal of Marketing Research}{46}{4}{543--556}.
\PrintBackRefs{\CurrentBib}

\end{thebibliography}

\end{document}